\documentclass[xelatex,prc,twocolumn,superscriptaddress,nofootinbib]{revtex4-1}

\usepackage{newtxtext}
\usepackage[varvw,bigdelims]{newtxmath}

\usepackage[colorlinks=true,allcolors=blue]{hyperref}
\usepackage{graphicx}
\usepackage{bm}
\usepackage{amsmath}
\usepackage{cancel}

\usepackage{physics}
\usepackage{xcolor}
\usepackage{orcidlink}

\newcommand{\mev}{\textrm{ MeV}}

\begin{document}
\title{ \boldmath On the molecular nature of the $\Omega_c(3120)$ and its analogy with the $\Omega(2012)$}

 \author{Natsumi Ikeno}
 \email{ikeno@tottori-u.ac.jp}
 \affiliation{Department of Agricultural, Life and Environmental Sciences, Tottori University, Tottori 680-8551, Japan}

\author{Wei-Hong Liang}%
\email{liangwh@gxnu.edu.cn}
\affiliation{Department of Physics, Guangxi Normal University, Guilin 541004, China}%
\affiliation{Guangxi Key Laboratory of Nuclear Physics and Technology, Guangxi Normal University, Guilin 541004, China}%


 \author{Eulogio Oset}
 \email{oset@ific.uv.es}
 \affiliation{Department of Physics, Guangxi Normal University, Guilin 541004, China}%
 \affiliation{Departamento de F\'{i}sica Te\'{o}rica and IFIC, Centro Mixto Universidad de Valencia - CSIC,
 Institutos de Investigaci\'{o}n de Paterna, Aptdo. 22085, 46071 Valencia, Spain}

\preprint{}

\date{\today}

\begin{abstract}
We make a study of the $\Omega_c(3120)$, one of the five $\Omega_c$ states observed by the LHCb collaboration, which is well reproduced as a molecular state from the $\Xi^*_c \bar K$ and $\Omega^*_c \eta$ channels mostly. The state with $J^P = 3/2^-$ decays to $\Xi_c \bar K$ in $D$-wave and we include this decay channel in our approach, as well as the effect of the $\Xi^*_c$ width. With all these ingredients, we determine the fraction of the $\Omega_c(3120)$ width that goes into $\Xi_c \pi \bar K$, which could be a measure of the $\Xi^*_c \bar K$ molecular component, but due to a relatively big binding, compared to its analogous $\Omega(2012)$ state, we find only a small fraction of about 3\%, which makes this measurement difficult with present statistics. As an alternative, we evaluate the scattering length and effective range of the $\Xi^*_c \bar K$ and $\Omega^*_c \eta$ channels which together with the binding and width of the $\Omega_c(3120)$ state, could give us an answer to the issue of the compositeness of this state when these magnitudes are determined experimentally, something feasible nowadays, for instance, measuring correlation functions.

\end{abstract}


\maketitle

\section{Introduction}
The $\Omega(2012)$ was reported by the Belle collaboration in Ref.~\cite{Belle:2018mqs} and estimulated much work from the theoretical side, 
some of it from the quark model perspective, assuming it to be a low-lying $p$-wave exited $3/2^-$state~\cite{Xiao:2018pwe,Aliev:2018yjo,Aliev:2018syi,Polyakov:2018mow,Liu:2019wdr,Arifi:2022ntc,Wang:2022zja}, or from the molecular picture generated by the interaction of the $\bar K \Xi^*(1530)$ and $\eta \Omega$ channels, decaying to $\bar K \Xi$~\cite{Valderrama:2018bmv,Lin:2018nqd,Pavao:2018xub,Huang:2018wth,Lu:2020ste,Ikeno:2020vqv,Liu:2020yen}.
The molecular picture is reinforced by the fact that the state was predicted before its experimental observation in Refs.~\cite{Hofmann:2006qx,Sarkar:2004jh}. It also gets extra support since even using quark models a molecular structure was claimed in Refs.~\cite{Wang:2007bf,Wang:2008zzz}. The use of the Weinberg compositeness condition also led the authors of Ref.~\cite{Gutsche:2019eoh} to advocate the molecular character of the state.

In order to test the molecular nature of the $\Omega(2012)$ the Belle collaboration conducted some tests, particularly looking at the decay into $\bar K \pi \Xi$, a signal of the $\bar K \Xi^*(1530)$ component of the state. A first experiment~\cite{Belle:2019zco} reported a ratio smaller than 11.9\% for the decay rate into $\bar K \pi \Xi$ versus $\bar K \Xi$, which might challenge the molecular picture, although not necessarily, as explained in Refs.~\cite{Lu:2020ste,Ikeno:2020vqv}. However, a posterior experiment~\cite{Belle:2022mrg} corrected this ratio and provided
\begin{equation}
   \mathcal{R}^{\Xi \pi \bar K}_{\Xi \bar K} = 0.97 \pm 0.24 \pm 0.07,
\end{equation}
and it was concluded that this ratio is consistent with the molecular interpretation of the $\Omega(2012)$ given in Refs.~\cite{Valderrama:2018bmv,Pavao:2018xub,Huang:2018wth,Gutsche:2019eoh}.

The basic idea on the molecular picture is that the $\Omega(2012)$ is a particular case of the interaction of the octet of pseudoscalar mesons with the baryons of the decuplet of the $\Delta(1232)$~\cite{Hofmann:2006qx,Sarkar:2004jh}.

Now we give a jump to the $\Omega_c$ states discovered by the LHCb collaboration~\cite{LHCb:2017uwr}. In this work five states were reported, $\Omega_c(3000)$, $\Omega_c(3050)$, $\Omega_c(3066)$, $\Omega_c(3090)$, $\Omega_c(3119)$. More recently two additional states have been found, $\Omega_c(3185)$ and $\Omega_c(3327)$~\cite{omcnew}.
The $\Omega_c$ states have also raised a wave of interest in the theoretical community, and, actually, many predictions about them and related states had been done, some from the quark model point of view \cite{Ebert:2007nw,Roberts:2007ni,Garcilazo:2007eh,Migura:2006ep,Ebert:2011kk,Valcarce:2008dr,Shah:2016nxi,Vijande:2012mk,Yoshida:2015tia,Chen:2015kpa,Chen:2016phw,Chiladze:1997ev,Manohar:1983md,Agaev:2017lip,Agaev:2017jyt,Wang:2017zjw}, and others from the molecular perspective~\cite{Hofmann:2005sw,Jimenez-Tejero:2009cyn,Romanets:2012hm,Xin:2023gkf}.
After the experimental discovery, work followed with several works trying to explain the states from the quark model perspective~\cite{Karliner:2017kfm,Wang:2017hej,Wang:2017vnc,Chen:2017gnu}, pentaquark structures~\cite{Yang:2017rpg,Huang:2017dwn,Kim:2017jpx,An:2017lwg,Ali:2017jda,Anisovich:2017aqa}, lattice QCD~\cite{Padmanath:2017lng}.

We follow here the molecular line and recall two independent works on the issue, an update of Ref.~\cite{Jimenez-Tejero:2009cyn} done in Ref.~\cite{Montana:2017kjw} to the light of the experimental results~\cite{omcnew} and the work of Ref.~\cite{Debastiani:2017ewu}.
Both of them use as input for the interaction the exchange of vector mesons based on the local hidden gauge approach~\cite{Bando:1984ej,Bando:1987br,Meissner:1987ge,Nagahiro:2008cv} between several coupled channels. There is only one free parameter, a cut off to regularize the loop functions, which is adjusted to get the mass of one state. 
The mass of the other states and the widths are then genuine prediction of the models. 
There is one difference between these works. 
In Ref.~\cite{Montana:2017kjw} SU(4) symmetry is used to obtain the baryon wave functions, while in Ref.~\cite{Debastiani:2017ewu} the wave functions involving $c$ quark are taken, as in Refs.~\cite{Capstick:1986ter,Roberts:2007ni}, isolating the heavy quarks and imposing the symmetry of the wave function on the light quarks. 
In Ref.~\cite{Montana:2017kjw} two $\Omega_c$ states were reproduced using as coupled channels pseudoscalar meson-baryon $1/2^+$ states, the $\Omega_c(3050)$ and $\Omega_c(3090)$ with $J^P = 1/2^-$. 
In Ref.~\cite{Debastiani:2017ewu} the same states were obtained, with practically the same properties, but in addition an extra state was obtained, the $\Omega(3120)$, with $J^P= 3/2^-$, coming from the interaction of pseudoscalar mesons with baryons of $3/2^+$, concretely the $\Xi^*_c \bar K$, $\Omega^*_c \eta$, and $\Xi^* D$, which were not considered in Ref.~\cite{Montana:2017kjw}. 
The agreement of the results of Refs.~\cite{Montana:2017kjw} and \cite{Debastiani:2017ewu} for the two states $\Omega_c(3050)$ and $\Omega_c(3090)$ is not accidental. Even if SU(4) symmetry is used in Ref.~\cite{Montana:2017kjw}, the important part of the interaction comes from the exchange of light vector mesons in which case the heavy quarks act as spectators and one is in practice projecting over SU(3), and the two pictures coincide. 
The two works of Refs.~\cite{Montana:2017kjw,Debastiani:2017ewu} conclude that the $\Omega_c(3120)$ state is not obtained as a $1/2^-$ state, but in Ref.~\cite{Debastiani:2017ewu} the state is obtained as a $3/2^-$ state. 
As mentioned above, the coupled channels in this case are $\Xi^*_c \bar K$, $\Omega_c^* \eta$, and $\Xi^* D$, with threshold masses $3142 \mev$, $3314 \mev$ and $3327 \mev$, respectively. 
The analogy with the $\Omega(2012)$ is clear. 
In this latter case, the coupled channels are $\Xi^* \bar K$, $\Omega \eta$. 
There is one extra channel, $\Xi^* D$, in the case of $\Omega_c(3120)$, but this channel plays a minor role in Ref.~\cite{Debastiani:2017ewu}. 
Indeed, the transition potential from $\Xi^* D$ to $\Xi_c^* \bar K$, $\Omega_c^* \eta$ is suppressed compared to that of $\Omega_c^* \eta$ to $\Xi^*_c \bar K$, the mass of the channel is more than $200 \mev$ above the mass of the $\Omega_c(3120)$ and the $\Omega_c(3120)$ wave function at the origin in coordinate space is almost $20$ times smaller than that of the $\Xi^*_c \bar K$ channel. 
We shall neglect this channel in our study, knowing that its small effect can be incorporated by small changes in the cut off parameter, which is fitted to the data. 
Thus, the analogy of the $\Omega(2012)$ and $\Omega_c(3120)$ is more apparent. There is also another common feature: in both cases, the state is not observed in any of the building blocks, instead the $\Omega(2012)$ is observed in the $\Xi \bar K$ channel and the $\Omega_c(3120)$ in the $\Xi_c \bar K$ one. 
Both channels appear in $D$-wave, they have not much relevance in the structure of the $\Omega(2012)$ and $\Omega_c(3120)$ states, but they provide the largest source of the width of the states, which is quite small, mostly due to the $D$-wave character of the decay.

In the present work, we retake the case of the $\Omega_c(3120)$ state and, by analogy to what was done in Ref.~\cite{Pavao:2018xub}, we introduce the $\Xi_c \bar K$ channel in $D$-wave in addition to the $\Xi^*_c \bar K$ and $\Omega^*_c \eta$ channels in $s$-wave, conduct a fit to the mass and width of the $\Omega_c (3120)$ state and evaluate the partial decay widths into $\Xi_c \bar K$ and $\pi \Xi_c \bar K$. The width was zero in Ref.~\cite{Debastiani:2017ewu} since the $\Xi_c \bar K$ decay channel was not included and the width of the $\Xi^*_c$ was also omitted.
At the same time, we evaluate the molecular probabilities of $\Xi_c^* \bar K$ and $\Omega^*_c \eta$ and find about 63\% and 10\% respectively, similar to those of the $\Xi^* \bar K$ and $\Omega \eta$ channels in the $\Omega(2012)$, indicating a large molecular component of the $\Omega_c(3120)$ wave function. The partial decay width into $\pi \Xi_c \bar K$ is found small, of the order of 3\%, much smaller than the corresponding $\pi \Xi \bar K$ one in the $\Omega(2012)$ case, the reason being that the $\Xi^*_c \bar K$ is now more bound than the $\Xi^* \bar K$ in the case of the $\Omega(2012)$ and the width of the $\Xi^*_c$ is much smaller than the binding energy of the $\Xi_c^* \bar K$ component. In order to find in experimental confirmation for the molecular structure of the $\Omega_c(3120)$ state, we evaluate the scattering length and effective range of the $\Xi^*_c \bar K$ and $\Omega_c^* \eta$ channels, which can be accessible in the future measuring correlation functions, and recall the works of Refs.~\cite{jingdai,jing1,jing2,haipeng} where it is found that the knowledge of the binding, scattering length, and effective range of the coupled channels that build up a molecular state can determine with a fair accuracy the molecular probability of the state.

\section{Formalism}

We take the results from Ref.~\cite{Debastiani:2017ewu} for the transition potential between the $\Xi_c^* \bar K$ and $\Omega^*_c \eta$ channels and introduce the $\Xi_c \bar K$ $D$-wave channel phenomenologically, as done in Ref.~\cite{Pavao:2018xub}.
The potential is given by
\begin{equation}\label{eq:Vmatrix}
  V=
  \begin{matrix}
    \begin{matrix}
      ~\Xi_c^* \bar K &~~~\Omega^*_c \eta~~&~~~\Xi_c \bar K~~ 
    \end{matrix}&\\[2.5mm]
    \begin{pmatrix}
      F&\frac{4}{\sqrt{3}}\, F&~\alpha\, q^{\,2}_{\rm on}~~\\[2mm]
      \frac{4}{\sqrt{3}}\, F&0&\beta\, q^{\, 2}_{\rm on}\\[2mm]
      ~~\alpha\, q^{\,2}_{\rm on}~&~\beta\, q^{\, 2}_{\rm on}~&0\\[2.5mm]
    \end{pmatrix}&
    \begin{matrix}
      ~\Xi_c^* \bar K\\[2mm]
      ~\Omega^*_c \eta\\[2mm]
      ~\Xi_c \bar K\\[2mm]
  \end{matrix}
\end{matrix}
\end{equation}
with
\begin{equation}
	F= - \dfrac{1}{4f^2}\; (k^0 +k'^{\,0}); ~~~f=93 \; {\rm MeV},
\end{equation}
\begin{equation}
	q_{\rm on} = \dfrac{\lambda^{1/2}(s, m^2_{\bar K}, m^2_{\Xi_c})}{2\, \sqrt{s}},
\end{equation} 
and $k^0, k'^{\,0}$ the energies of the initial, final meson. In Eq.~\eqref{eq:Vmatrix} $\alpha, \beta$ are unknown parameters, to be fitted to the width of the $\Omega_c (3120)$ state.

The scattering matrix between these three channels is given by
\begin{equation}\label{eq:BS}
  T=[1-VG]^{-1} \, V,
\end{equation}
where $G$ is the diagonal matrix of loop function for the meson-baryon states, ${\rm diag} (G_{\Xi^*_c \bar K}, G_{\Omega^*_c \eta }, G_{\Xi_c \bar K})$, with 
\begin{eqnarray}\label{eq:Gi}
  G_i(\sqrt{s}) &=& \int_{|\vec q\,| < q_{\rm max}} \dfrac{{\rm d}^3 q}{(2\pi)^3} \; \dfrac{1}{2\,\omega_i(q)}\;\nonumber\\[2mm]
   &&\times\dfrac{M_i}{E_i(q)}\;\dfrac{1}{\sqrt{s}-\omega_i(q)-E_i(q) + i \varepsilon},
\end{eqnarray}
for the $S$-wave $\Xi_c^* \bar K$ and $\Omega^*_c \eta$ channel, and
\begin{eqnarray}\label{eq:G3}
  G_{\Xi_c \bar K}(\sqrt{s})&=& \int_{|\vec q\,| < q_{\rm max}} \dfrac{{\rm d}^3 q}{(2\pi)^3} \; \left( \dfrac{q}{q_{\rm on}}\right)^4 \, \dfrac{1}{2\,\omega_{\bar K}(q)}\nonumber\\[2mm]
  &&\times \dfrac{M_{\Xi_c}}{E_{\Xi_c}(q)}\;\dfrac{1}{\sqrt{s}-\omega_{\bar K}(q)-E_{\Xi_c}(q) + i \varepsilon},
\end{eqnarray}
for the $D$-wave $\Xi_c \bar K$ channel, where 
$\omega_i(q)=\sqrt{m_i^2+\vec q^{\;2}}$, $E_i(q)=\sqrt{M_i^2+\vec q^{\;2} }$, with $m_i, M_i$ the masses of meson and baryon in channel $i$.
For the parameter $q_{\rm max}$, we shall take a value around $650 \, \mev$, as in Ref.~\cite{Debastiani:2017ewu}, fine tuned to get the right energy of the state. The parameters $\alpha,\beta$ will be chosen to get the width of the $\Omega_c(3120)$ state. We have from Ref.~\cite{LHCb:2017uwr} \footnote{In Ref.~\cite{omcnew} the $\Omega_c(3120)$ width is changed to $0.60\pm 0.63 \mev$, which overlaps with the results of Eq.~\eqref{eq:LHCbOmec}. We carry our analysis using the data of Eq.~\eqref{eq:LHCbOmec}. The conclusions of the paper do not change from using one or the other data.},
\begin{equation}\label{eq:LHCbOmec}
  \begin{split}
  &M_{\Omega_c(3120)}  =  3119.1\pm 0.3 \pm 0.9^{+0.3}_{-0.5} \; \mev,  \\[2mm]
  &\Gamma_{\Omega_c(3120)}   =  1.1 \pm  0.8 \pm 0.4 \; \mev. 
\end{split} 
\end{equation}
  
  In order to see the relevance of the $\Xi_c^*$ decay width in the width of the $\Omega_c (3120)$ state, we calculate the $T$ matrix including the $\Xi_c^*$ selfenergy in the loop. For this we follow the prescription given in Ref.~\cite{Dai:2022ulk}, recommended when the width of the particle is small, instead of the popular convolution method used in Ref.~\cite{Pavao:2018xub}, substituting the $G$ function for the $\Xi_c^* \bar K$ channel by
  \begin{eqnarray}\label{eq:G1width}
    \tilde{G}_{\Xi^*_c \bar K}(\sqrt{s})&=& \int_{|\vec q\,| < q_{\rm max}} \dfrac{{\rm d}^3 q}{(2\pi)^3} \; \dfrac{1}{2\,\omega_{\bar K}(q)}\; \dfrac{M_{\Xi_c^*}}{E_{\Xi_c^*}(q)} \nonumber\\[2mm]
    &\times& \dfrac{1}{\sqrt{s}-\omega_{\bar K}(q)-E_{\Xi_c^*}(q) + i \frac{\sqrt{s'}}{2 E_{\Xi_c^*} (q)}  \Gamma_{\Xi_c^{*}}(\sqrt{s'})},\nonumber\\
  \end{eqnarray}
  where
  \begin{equation}
    s'=[\sqrt{s}-\omega_{\bar K}(q) ]^2-\vec q^{\,2},
  \end{equation}
  and 
  \begin{equation}\label{eq:GamXic}
    \Gamma_{\Xi_c^{*}}(M_{\rm inv})= \dfrac{M_{\Xi_c^*}}{M_{\rm inv}}\, \left( \dfrac{q'}{q'_{\rm on}} \right)^3 \, \Gamma_{\rm on} \; \theta (M_{\rm inv}-m_\pi -M_{\Xi_c}),
  \end{equation}
  with $M_{\rm inv} = \sqrt{s'}$ the invariant mass of $\pi \Xi_c$, $\Gamma_{\rm on}=2.25 \mev$ the average width of $\Xi_c^{*+}$ and $\Xi_c^{*0}$ from PDG, and 
  \begin{equation}\label{eq:qon}
    \begin{split}
      q'_{\rm on}&=\dfrac{\lambda^{1/2}(M^2_{\Xi_c^*}, m^2_\pi, M^2_{\Xi_c})}{2\, M_{\Xi_c^*}}, \\[2mm]
      q'&=\dfrac{\lambda^{1/2}(M^2_{\rm inv}, m^2_\pi, M^2_{\Xi_c})}{2\, M_{\rm inv}}, 
  \end{split} 
  \end{equation}

  We also need to evaluate the couplings at the pole. For that we must use $G$ in the second Riemann sheet
\begin{equation}\label{eq:GII}
    G^{(II)}_i (\sqrt{s})= G_i(\sqrt{s})+ 
    \left\{
          \begin{array}{ll}
                 0, & ~~{\rm for} \; {\rm Re}(\sqrt{s}) < \sqrt{s}_{{\rm th}, i}  \\[2mm]
                 i \dfrac{M_i \,k}{2\pi \sqrt{s}}, &~~{\rm for} \; {\rm Re}(\sqrt{s})  \geqslant  \sqrt{s}_{{\rm th}, i},
          \end{array}
    \right.
    \end{equation}
    with $\sqrt{s}_{{\rm th}, i}$ the threshold mass of channel $i$, and   
    \begin{equation}\label{eq:k}
      k=\dfrac{\lambda^{1/2}(s, m_i^2, M_i^2)}{2\sqrt{s}}, ~~~~ {\rm Im}(k)>0.
      \end{equation}
Only the $\Xi_c \bar K$ channel goes to the second Riemann sheet at the pole.

The couplings are defined at the pole as 
\begin{equation}\label{eq:gigj}
  g_i g_j=\lim_{\sqrt{s} \to \sqrt{s_p}} (\sqrt{s}-\sqrt{s_p})\;T_{ij},
  \end{equation}
with $\sqrt{s_{\rm p}}$ the energy of the pole, which allows to have the relative phase of one coupling to another, with one of them chosen with arbitrary phase. We choose $g_{\Xi_c^* \bar K}$ positive. We evaluate the couplings neglecting the width of the $\Xi_c^*$.

\begin{figure}[t]
  \begin{center}
  \includegraphics[scale=0.47]{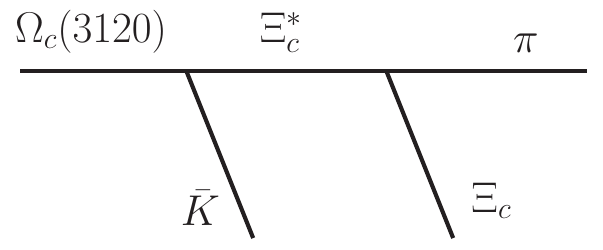}
  \end{center}
  \vspace{-0.7cm}
  \caption{Mechanism for $\Omega_c (3120)$ to decay into $\Xi_c \bar K \pi$ via primary decay into $\Xi_c^* \bar K$.}
  \label{fig:Fig1}
  \end{figure}

Once the couplings are evaluated we can calculate the molecular probabilities of the $S$-wave channels as
\begin{equation}\label{eq:Pi}
  \mathcal{P}_i=- g_i^2 \; \dfrac{\partial G_i}{\partial \sqrt{s}}.
\end{equation}
Finally, in order to evaluate the partial decay width into the $\Xi_c \bar K \pi$ channel, in analogy to Ref.~\cite{Ikeno:2020vqv}, we evaluate the amplitude for the diagram of Fig. \ref{fig:Fig1} as
\begin{equation}\label{eq:t3}
  t_{\Omega_c \to \pi \bar K \Xi_c}= g_{\Omega_c, \bar K \Xi^*_c} \;\dfrac{1}{M_{\rm inv}(\pi \Xi_c)-M_{\Xi_c^*}-i\Gamma_{\Xi_c^*}/2}\; g_{\Xi_c^*, \pi \Xi_c} \; \tilde{p}_\pi,
\end{equation}
where
\begin{equation}\label{eq:ppi}
  \tilde{p}_\pi = \dfrac{\lambda^{1/2}(M^2_{\rm inv}(\pi \Xi_c), m_\pi^2, M^2_{\Xi_c})}{2\, M_{\rm inv}(\pi \Xi_c)},
\end{equation}
with $g_{\Xi_c^*, \pi \Xi_c}$ given using $\Gamma_{\Xi_c^*}$ as 
\begin{equation}\label{eq:Gamc}
  \Gamma_{\Xi_c^*}= \dfrac{1}{2\,\pi}\;\dfrac{M_{\Xi_c}}{M_{\Xi_c^*}}\; g^2_{\Xi_c^*, \pi \Xi_c} \,\tilde{p}^3_\pi,
\end{equation}
with $\Gamma_{\Xi_c^*}$ given by Eq.~\eqref{eq:GamXic}.

The mass distribution for the $3$ body decay of the mechanism of Fig. \ref{fig:Fig1} is given by
\begin{equation}\label{eq:Minv}
  \dfrac{{\rm d} \Gamma_{\Omega_c}}{{\rm d} M_{\rm inv}(\pi \Xi_c)}= \dfrac{1}{(2\pi)^3}\; \dfrac{M_{\Xi_c}}{M_{\Omega_c}}\; p_{\bar K} \tilde{p}_\pi\; |t_{\Omega_c \to \pi \bar K \Xi_c}|^2,
\end{equation}
with 
\begin{equation}\label{eq:pk}
  p_{\bar K}= \dfrac{\lambda^{1/2}(M_{\Omega_c}^2, m^2_{\bar K}, M^2_{\rm inv}(\pi \Xi_c))}{2\, M_{\Omega_c}},
\end{equation}
and the integration over $M_{\rm inv}(\pi \Xi_c)$ produces the width for $\Omega_c (3120)$ decay to $\pi \bar K \Xi_c$.

The width for $\Omega_c (3120) \to \Xi_c \bar K $ decay can be obtained from 
\begin{equation}\label{eq:Gam2body}
  \Gamma= \dfrac{1}{2\pi}\; \dfrac{M_{\Xi_c}}{M_{\Omega_c}}\; g^2_{\Omega_c, \Xi_c \bar K} \; p'_{\bar K},
\end{equation}
 with $p'_{\bar K}$ the $\bar K$ momentum for $\Omega_c \to \Xi_c \bar K $ decay in the $\Omega_c$ rest frame.

 \section{Results}
 We conduct a fit to the position and width of the $\Omega_c (3120)$ state using Eq.~\eqref{eq:BS} with the potential of Eq.~\eqref{eq:Vmatrix}, the $G$ functions of Eqs.~\eqref{eq:Gi} and \eqref{eq:G3} using the second Riemann sheet of Eq.~\eqref{eq:GII}. We get a good fit to the data with the parameters 
  \begin{equation}\label{eq:param}
  \begin{split}
    q_{\rm max}&= 674.6 \, \mev, \\
    \alpha&= 2.6 \times 10^{-8} \, \mev^{-3}, \\
    \beta&= 2.0 \times 10^{-9} \, \mev^{-3}.
\end{split} 
\end{equation}
The pole position appears at 
 \begin{equation}\label{eq:pole}
   (3119.13 +i \,0.54)\, \mev
 \end{equation}
 implying a width of $1.08 \mev$ in agreement with the central values of the experiment.
 
\begin{figure}[b]
   \begin{center}
   \includegraphics[scale=0.48]{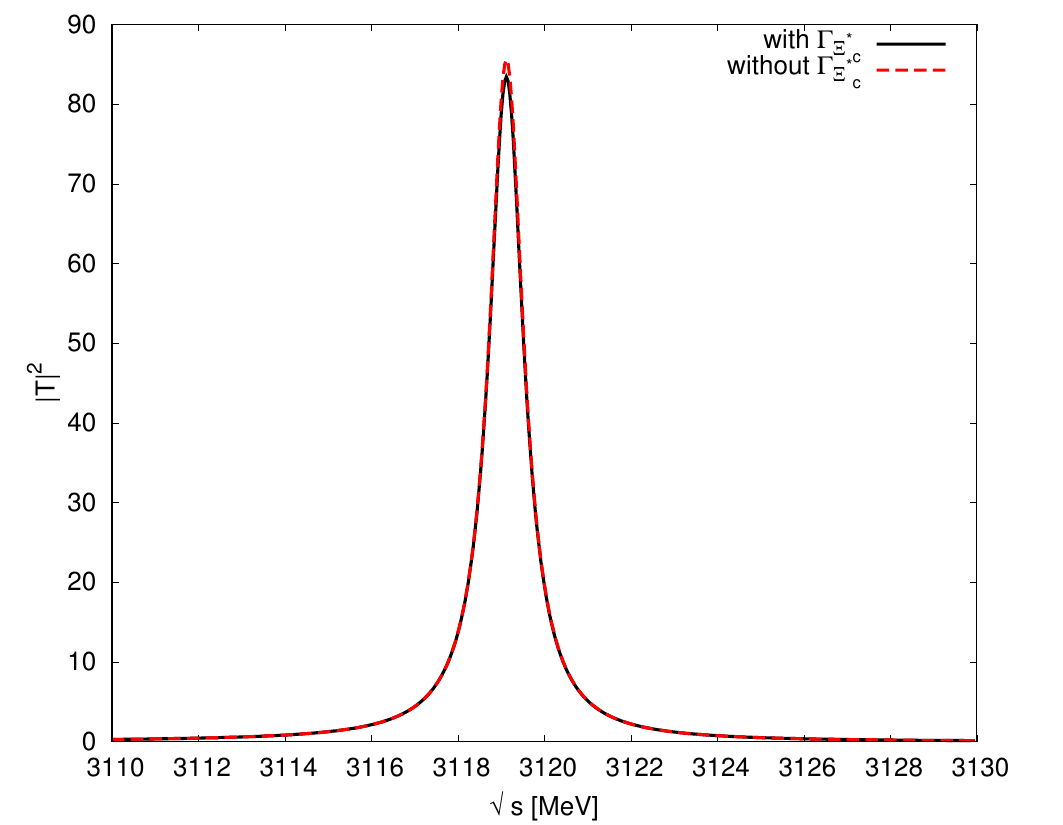}
   \end{center}
   \vspace{-0.7cm}
   \caption{$|T_{\Xi_c^* \bar K}|^2$ as a function of $\sqrt{s}$ in the cases with $\Gamma_{\Xi_c^*}$ and without $\Gamma_{\Xi_c^*}$, respectively.}
   \label{fig:Fig2}
   \end{figure}

 Now we add the $\Xi_c^*$ selfenergy in Eq.~\eqref{eq:G1width} and look at the $T_{\Xi_c^* \bar K}$ amplitude in the real energy axis. In Fig.~\ref{fig:Fig2}
 we show the results for $|T_{\Xi_c^* \bar K}|^2$ with and without the $\Xi_c^*$ width. As we see, the results are very similar, with the masses and widths practically the same. This means that, unlike in the case of the $\Omega(2012)$, where the difference between the widths in the analogous cases, with $\Xi^*$ dressed or without, allowed us to determine the decay width of $\Omega(2012) \to \Xi^* \bar K \to \Xi \pi \bar K$, in the present case we cannot determine the $\Omega_c(3120) \to \Xi_c^* \bar K \to \Xi_c \pi \bar K$ with precision using this procedure. Hence, we use the more accurate one of evaluating the width using explicitly the mechanism of Fig.~\ref{fig:Fig1}.
 For this we need the coupling of $\Omega_c(3120) \to \Xi_c^* \bar K$, which we address below. The couplings are evaluated at the pole using the $G$ functions in the second Riemann sheet through Eq.~\eqref{eq:GII} and we find the results of Table \ref{tab:gigG}.
 \begin{table}[t]
 \renewcommand{\arraystretch}{0.9}
 \setlength{\tabcolsep}{0.3cm}
 \centering
 \caption{The couplings $g_i$ of $\Omega_c(3120)$ to different channels, $g_i G_i$ and the molecular probabilities for the $S$-wave channels.}
 \label{tab:gigG}
 \begin{tabular}{c|ccc}
 \hline
  & $\Xi_c^* \bar K$  & $\Omega^*_c \eta$ & $\Xi_c \bar K$ \\[3mm]
  \hline
   $g_i$  & $2.06 -i 0.02$ & $2.09-i0.01$  & $-0.138$ \\[3mm]
   $g_i G_i$ & $-36.77 +i0.17$  & $-17.64+i 0.06$ & \\[3mm]
    $-g_i^2 \; \dfrac{\partial G_i}{\partial \sqrt{s}}$ & $0.63$ & $0.10$  &   \\[4mm]
 \hline\hline
 \end{tabular}
 \end{table}
 We also show there the values of $g_i G_i$ for the $S$-wave channels, which are the wave functions at the origin in coordinate space \cite{Gamermann:2009uq}. 
As we can see, the wave function is dominated by the $\Xi_c^* \bar K$ component. We ignore now the tiny imaginary part of the couplings and calculate the probability of the $\Xi_c^* \bar K, \Omega^*_c \eta$ channels through $-g_i^2 \; \dfrac{\partial G_i}{\partial \sqrt{s}}$ \cite{Gamermann:2009uq, Hyodo:2013nka}. 
We see again that the $\Xi_c^* \bar K$ has the largest probability of around $63\%$ and the $\Omega_c^* \eta$ around $10 \%$, hence we have a largely molecular state.
 
 Once we have calculated the couplings we are in a position to evaluate the $\Omega_c(3120) \to \Xi_c^* \bar K \to \Xi_c \pi \bar K$ for the mechanism of Fig.~\ref{fig:Fig1} through Eqs.~\eqref{eq:t3}, \eqref{eq:Gamc} and \eqref{eq:Minv}.
 We get 
 \begin{equation}
   \Gamma_{\Omega_c \to \Xi_c \pi \bar K} =0.03  \mev,
 \end{equation}
 while through Eq.~\eqref{eq:Gam2body} we would get 
 \begin{equation}
   \Gamma_{\Omega_c \to \Xi_c \bar K} =0.90  \mev,
 \end{equation}
 the sum of them giving $\Gamma_{\Omega_c} \sim 1 \mev$ as the central value of the experiment, Eq.~\eqref{eq:LHCbOmec}.
 The fraction of $\Omega_c (3120)$ decay to $\Xi_c \pi \bar K$ is much smaller than the $\Xi \pi \bar K$ in the case of the $\Omega(2012)$.
 However, the molecular probabilities of the $\Xi^*_c \bar K$ and $\Xi^* \bar K$ in both cases are very similar. The differences stem from the different bindings. In the case of the $\Omega(2012)$ the diagonal terms in the $V$ matrix of Eq.~\eqref{eq:Vmatrix} were zero, while here they are finite and negative, indicating extra attraction that reverts into a much bigger binding of about $20 \mev$. This has as a consequence that the decay of the $\Xi_c^* \bar K$ bound component into $\Xi^*_c \bar K \to \Xi_c \pi \bar K$ is more difficult (technically, the $\Xi_c^*$ in the diagram of Fig.~\ref{fig:Fig1} is more off shell than the $\Xi^*$ in the analogous diagram for $\Omega(2012)$ decay).
 While the experimental determination of a fraction of $0.03 \mev$ is certainly challenging given the present experimental errors in Eq.~\eqref{eq:LHCbOmec}, we look into other experimental tests that can lead us to determine the nature of that state. In Refs.~\cite{jingdai,jing1,jing2,haipeng} it was discussed in detail how the knowledge of the binding, scattering length and effective range of the coupled channels provided excellent information on the molecular compositeness of the states. The formalism, improved substantially over the possible application of Weinberg formulas \cite{Weinberg:1965zz}, where that range is ignored and leads to unrealistic results in most cases.
Anticipating that these magnitudes can be determined, for instance, using correlation functions \cite{Tolos:2020aln}, we determine here the scattering length and effective range of the $\Xi_c^* \bar K$ and $\Omega_c^* \eta$ channels using the formulas of Ref.~\cite{Molina:2023jov},
 \begin{equation}\label{eq:aj2}
   -\dfrac{1}{a_j}=\left. -\dfrac{8\pi \,\sqrt{s}}{2\, M_j}\; (T_{jj})^{-1}\right|_{\sqrt{s_{{\rm th}}}, j},
   \end{equation}
   \begin{equation}\label{eq:r0j}
   r_{0, j}=
    \dfrac{1}{\mu_j}\; \dfrac{\partial}{\partial \sqrt{s}}\,
   \left[ -\dfrac{8\pi \,\sqrt{s}}{2\, M_j}\; (T_{jj})^{-1}+ik_j \right]_{\sqrt{s_{{\rm th}}}, j},
   \end{equation}
with $\mu_j$ the reduced mass of channel $j$ and $k_j$ the momentum of a particle of the pair in their rest frame. 
The results obtained are shown in Table \ref{tab:ar0}.
 \begin{table}[t]
 \renewcommand{\arraystretch}{0.9}
 \setlength{\tabcolsep}{0.3cm}
 \centering
 \caption{Scattering length $a_j$ and effective range $r_{0, j}$ of the $\Xi_c^* \bar K$ and $\Omega_c^* \eta$ channels. [in units of fm]}
 \label{tab:ar0}
 \begin{tabular}{c|ccc}
 \hline
  & $a_j$  & $r_{0, j}$   \\[3mm]
  \hline
  $\Xi_c^* \bar K$  & $1.45 -i 0.07$ & $-0.08-i0.01$   \\[3mm]
  $\Omega^*_c \eta$ & $0.44 -i0.09$  & $0.26+i 0.01$ \\
 \hline\hline
 \end{tabular}
 \end{table}
 We can see that the values of $a_j$ and $r_{0, j}$ are mostly real, because the main source of decay is to $\Xi_c \bar K$, which we saw leads to a very small width of the $\Omega_c(3120)$ state. 
 
 As we can see, we are providing new magnitudes, which are additional to the binding and width of the $\Omega_c(3120)$ state. We believe that these magnitudes should be sufficient to determine the nature of the $\Omega_c(3120)$ state and the precent work should give an incentive to do experimental work in this direction.
 
 \section{Conclusions}
 We made a thorough study of the $\Omega_c(3120)$ state, reported in the LHCb experiment in Ref.~~\cite{LHCb:2017uwr}, which was shown in Ref.~~\cite{Debastiani:2017ewu} to be well reproduced as a molecular state from the interaction of mostly the $\Xi^*_c \bar K$ and $\Omega_c^* \eta$ channels. The state has $J^P= \frac{3}{2}^-$ and is distinct from $\Omega_c (3050), \Omega_c (3090) (\frac{1}{2}^-)$ which are generated from the pseudoscalar-baryon $(\frac{1}{2}^+)$ interaction in Refs.~~\cite{Montana:2017kjw,Debastiani:2017ewu}.
  In addition to the $\Xi^*_c \bar K, \Omega_c^* \eta$ channels used in Ref.~\cite{Debastiani:2017ewu}, we include now the $\Xi_c \bar K$ decay channel, where the state was observed, which appears in $D$-wave.
 With the consideration of the $\Xi_c \bar K$ channel, we can now obtain the width of the state, which was not evaluated in Ref.~~\cite{Debastiani:2017ewu}.
Then we also evaluate the $\Xi_c \pi \bar K$ decay width, which could be a measure of the $\Xi^*_c \bar K$ component of the $\Omega_c(3120)$ state, but unlike in the analogous $\Omega(2012)$ state where the $\Xi \pi \bar K$ decay channel is sizeable and has been measured, providing support for the $\Xi^* \bar K$ molecular picture of the $\Omega(2012)$ state, here we obtain a very small fraction of $\Xi_c \pi \bar K$, because the $\Omega_c(3120)$ state is more bound than the $\Omega(2012)$. In view of this, we determined the scattering length and effective range of the $\Xi^*_c \bar K$ and $\Omega_c^* \eta$ channels and made a call for the experimental determination of these magnitudes, accessible for instance measuring correlation functions, which can be instrumental in the determination of the nature of the $\Omega_c(3120)$ state.


\section*{Acknowledgments}
One of us, N. I. wishes to acknowledge the hospitality of Guangxi Normal University, where part of the work was carried out.
This work is partly supported by the National Natural Science Foundation of China under Grant No. 11975083 and No. 12365019, and by the Central Government Guidance Funds for Local Scientific and Technological Development, China (No. Guike ZY22096024).
This work is also partly supported by the Spanish Ministerio de Economia y Competitividad (MINECO) and European FEDER
funds under Contracts No. FIS2017-84038-C2-1-P B, PID2020-112777GB-I00, and by Generalitat Valenciana under contract
PROMETEO/2020/023.
This project has received funding from the European Union Horizon 2020 research and innovation
programme under the program H2020-INFRAIA-2018-1, grant agreement No. 824093 of the STRONG-2020 project.




%

\end{document}